Towards Generalizable Drowsiness Monitoring with Physiological Sensors: A Preliminary Study

Jiyao Wang[1], Suzan Ayas[2], Jiahao Zhang[1], Xiao Wen[3], Dengbo He[1], Birsen Donmez[2]

[1] The Hong Kong University of Science and Technology (Guangzhou)

[2] University of Toronto

[3] The Hong Kong University of Science and Technology

Accurately detecting drowsiness is vital to driving safety. Among all measures, physiological-signal-based drowsiness monitoring can be more privacy-preserving than a camera-based approach. However, conflicts exist regarding how physiological metrics are associated with different drowsiness labels across datasets. Thus, we analyzed key features from electrocardiograms (ECG), electrodermal activity (EDA), and respiratory (RESP) signals across four datasets, where different drowsiness inducers (such as fatigue and low arousal) and assessment methods (subjective vs. objective) were used. Binary logistic regression models were built to identify the physiological metrics that are associated with drowsiness. Findings indicate that distinct different drowsiness inducers can lead to different physiological responses, and objective assessments were more sensitive than subjective ones in detecting drowsiness. Further, the increased heart rate stability, reduced respiratory amplitude, and decreased tonic EDA are robustly associated with increased drowsiness. The results enhance understanding of drowsiness detection and can inform future generalizable monitoring designs.

## INTRODUCTION

Drowsiness in driving is defined as a state of impaired consciousness where a driver is more inclined to sleep than to stay awake (Slater, 2008), which can lead to fatal consequences. For example, an estimated 328,000 crashes involve drowsy drivers in the US each year, resulting in 109,000 injuries and 6,400 fatalities (Tefft et al., 2012). Drowsiness can lead to decreased alertness, reaction times, and decision-making abilities (Ashraf et al., 2019). As technology advances, data-driven methods have emerged to detect drowsiness using images, physiological signals, and driving behaviors (Saleem et al., 2023). Particularly, due to infrequent manual driving in automated vehicles and privacy concerns, there is a growing interest in monitoring drowsiness through physiological signals (Kakhi et al., 2024). Physiological sensors that are commonly used for drowsiness monitoring in the in-vehicle environment include electroencephalography (EEG), electrocardiogram (ECG), electrodermal activity (EDA), and respiration (RESP). Among them, EEG, while more accurate, is practically infeasible at this stage in the real world due to its intrusiveness and high sensitivity to motion noise (Saleem et al., 2023). Therefore, ECG, RESP, and EDA are commonly used in previous research.

Although data-driven methods based on physiological measures have shown promising results in detecting driver drowsiness, their effectiveness in real-world applications remains limited, partly due to the lack of high-quality training data, which is essential for effective drowsiness detection. Specifically, previous research usually took different approaches to induce drowsiness, such as by long-boring driving or sleep deprivation. However, drowsy drivers induced by different inducers, such as low arousal and sleep deprivation, might exhibit different physiological responses. Different inducers, such as low arousal and sleep deprivation, can lead to different physiological responses (Ayas et al., 2023). Given that the varied experimental conditions can limit the applicability of findings from one specific study to others (Mason et al., 2003), current drowsiness monitoring systems (Alguindigue et al., 2024; Meteier et al., 2024), trained on specific datasets, may struggle to generalize to varied causes of drowsiness in real-world scenarios.

Further, there is heterogeneity in how drowsiness was assessed across studies (e.g., self-report questionnaires, yawning). Various criteria were used to evaluate driver drowsiness, which can generally be categorized as: (1) Subjective assessments, such as self-assessment questionnaires and the Karolinska Sleepiness Scale (KSS) (Kaida et al., 2006), recognized as the driver drowsiness criterion by the European Union (EUR-Lex, 2021); (2) Objective indicators, such as eye-blinking and yawning, widely utilized in some standards in China (The Center for Automotive Assessment and Management, 2024). Although some researchers (Sakata et al., 2024) suggested replacing KSS with objective indicators, their suggestions are still based on one experiment with one specific drowsiness inducer and a limited participant size, which may affect the generalizability of their findings. Further, the concept of fatigue and drowsiness was often used interchangeably, while fatigue is just one of the inducers of drowsiness. As such, the inconsistency regarding how the drowsiness states are labelled in the existing dataset may also impair the generalizability of the association between the key physiological features and the "true" drowsiness states.

Thus, in this study, we aim to identify consistently robust physiological metrics of drowsiness by validating the drowsiness-related physiological features from four datasets, each using different approaches for inducing and assessing drowsiness. We have three research questions (RQs): **RQ1**. Do different contributing factors of drowsiness show variations in their association with key physiological features and the induced probability of being drowsy? **RQ2**. Are there variations in drowsiness recognition capabilities across different drowsiness assessment methods? **RQ3**. Which physiological indicators consistently correlate with different drowsiness types identified by different assessment methods? The findings aim to enhance generalizability for a large-scale training dataset and inform future drowsiness monitoring system designs.

Table 1. Descriptive Statistics of the Investigated Datasets.

| Dataset | Date | Subject | Age (Mean, Min-Max, SD) | Gender | Collection Device | Sampling Frequency | Drowsiness Criteria |
|---|---|---|---|---|---|---|---|
| Fatigueset | 2021 | 12 | 30.75, 21-40, 5.8 | 9 male; 3 female | BioHarness, E4 | 100 HZ | Objective |
| AdVitam | 2023 | 63 | 23.8, 18-64, 4.8 | 45 male; 18 female | BioPac MP36 | 1000 HZ | Subjective |
| MCDD | 2024 | 42 | 35.3, 23-53, 9.1 | 25 male; 17 female | Ergoneers | 100 HZ | Subjective |
| Ayas et. al | 2024 | 27 | 36.7, 19-74, 14.4 | 14 male; 14 female | Becker Meditec | 256 HZ | Objective |

## APPROACH

### Datasets

We analyzed four datasets—three publicly available and one private—whose key characteristics are summarized in Table 1. The Fatigueset dataset (Kalanadhabhatta et al., 2021) comprises recordings from 12 participants who performed physical exercise to induce fatigue; two-minute segments immediately before and after exercise are labelled "normal" and "fatigue," respectively. Although Fatigueset does not involve driving, we included it to broaden our coverage of drowsiness-inducing conditions.

The AdVitam dataset (Meteier et al., 2023) captures drowsiness in 63 drivers during SAE Level 3 automated driving. We focus on "Experiment 4," in which participants were sleep-deprived (under six hours of sleep the previous night) and completed one hour of on-road driving, with drowsiness assessed via the KSS questionnaire.

Similarly, the MCDD dataset (Wang et al., 2024) involves 42 drivers who, over 2.5 hours of L3 automated driving, performed three non-driving cognitive tasks (N-back, mental arithmetic, and spatial search); drowsiness ratings were collected every five minutes using the KSS.

Finally, Ayas et al. (2024) is private dataset. It observed 27 drivers in a monotonous Level 2 automated driving scenario lasting up to 1.5 hours without surrounding traffic interactions. Drowsiness annotations were provided by independent raters based on objective facial-video indicators (Wierwille & Ellsworth, 1994).

### Preprocessing

We first standardized all signals from all datasets to a sampling frequency of 100 Hz using linear interpolation and downsampling to ensure consistency. EDA was then smoothed with a low-pass filter at 5 Hz, while ECG and RESP signals were each band-pass filtered (3~45 Hz for ECG and 0.1~0.35 Hz for RESP) to remove noise outside their physiological ranges (Meteier et al., 2021). Then, following Meteier et al. (2024), we extracted 32 physiological features in all from each segment for further analysis. Note that, for the datasets labelled via the KSS, we binarized sleepiness according to established thresholds: any sample with KSS ≥ 7 was marked as 'drowsy', whereas samples with KSS < 6 were considered 'awake' (Hultman et al., 2021).

### Statistics Models

Five binary logistic regression models (i.e., Model (a) to (e)) were built in "SAS OnDemand for Academics" using the

GENMOD procedure, with drowsiness/fatigue (yes vs. no) as the dependent variable. A generalized estimation equation accounted for repeated measures on each participant. The first four models were based on different datasets, while the last model was built on the aggregation of four datasets.

Table 2. Features Calculated from Physiological Signals.

| Feature | Description |
|---|---|
| ECG-HR | The average number of heartbeats per minute. |
| ECG-RMSSD | The square root of the mean of successive differences between adjacent inter-beat intervals. |
| ECG-LF | The power in the low-frequency (0.04–0.15 Hz) band of heart rate variability. |
| ECG-LF/HF | The ratio of low-frequency to high-frequency HRV power, indicating autonomic balance. |
| RESP-RR | The average number of complete breaths per minute. |
| RESP-Mean Amplitude | The mean peak-to-trough amplitude of respiratory cycles. |
| RESP-RMSSD | The square root of the mean of successive differences between consecutive respiratory cycle durations. |
| RESP-Phase Duration Ratio | The ratio of inspiratory to expiratory phase durations. |
| EDA-Std tonic | The standard deviation of the tonic EDA signal. |
| EDA-Mean tonic | The mean value of the tonic EDA signal. |
| EDA-Min tonic | The minimum value of the tonic EDA signal. |
| EDA-Max tonic | The maximum value of the tonic EDA signal. |
| EDA-SCR Mean Amplitude | The average amplitude of skin conductance responses. |
| EDA-SCR Peak Num | The total number of detected skin conductance response peaks. |

To address multicollinearity issues, Pearson correlations (r) were calculated among 32 physiological features, with highly correlated variables (r > 0.8 or r < -0.8, $p < .05$) either aggregated or discarded based on the Quasi-likelihood under the Independence Model Criterion (QIC). Finally, 14 features were reserved as independent variables for all models, which are explained in Table 2, along with their descriptions. During the model construction process, all physiological features were included. Then, a backward stepwise selection was performed based on QIC to obtain the best-fitted models.

For the last Model (e), except for the 14 physiological features, we annotated samples from different datasets with one distinct ID and named this new independent variable as Drowsy Type, which reflected drowsiness/fatigue caused by different inducers. In the Model (e), the post-hoc comparisons were conducted for Drowsy Type if it was significant ($p < .05$).

## RESULTS

Table 3. Summary of Model (a)'s Results.

| Model a: Fatigueset (Physical Fatigue, Objective) | | | |
|---|---|---|---|
| Independent Variable | Estimate [95% CI] | $\chi^2$-value | p-value |
| ECG-HR | 0.14 [0.02, 0.26] | $\chi^2(1) = 5.38$ | .02** |
| ECG-LF/HF | -0.19 [-0.32, -0.06] | $\chi^2(1) = 8.23$ | .004** |
| RESP-RR | 0.46 [0.45, 0.88] | $\chi^2(1) = 4.72$ | .03** |
| EDA-Mean tonic | 54.28 [9.60, 98.96] | $\chi^2(1) = 5.67$ | .002** |
| EDA-Min tonic | 2.97 [1.32, 4.62] | $\chi^2(1) = 12.46$ | .0004** |
| EDA-SCR Mean Amplitude | 1479.6 [-131.71, 3090.91] | $\chi^2(1) = 3.24$ | .08* |

Table 4. Summary of Model (b)'s Results.

| Model b: Ayas et.al (Low Arousal, Objective) | | | |
|---|---|---|---|
| Independent Variable | Estimate [95% CI] | $\chi^2$-value | p-value |
| ECG-HR | 0.16 [0.12, 0.19] | $\chi^2(1) = 86.53$ | <.0001** |
| RESP-Mean Amplitude | 1.56 [0.94, 2.18] | $\chi^2(1) = 24.06$ | <.0001** |
| RESP-RR | -0.15 [-0.31, 0.0] | $\chi^2(1) = 3.94$ | .047** |
| EDA-Min tonic | -0.0002 [-0.0002, -0.0001] | $\chi^2(1) = 27.95$ | <.0001** |
| EDA-SCR Peak Num | 0.02 [0.01, 0.02] | $\chi^2(1) = 32.79$ | <.0001** |

Table 5. Summary of Model (c)'s Results.

| Model c: AdVitam (Sleep Deprivation, Subjective) | | | |
|---|---|---|---|
| Independent Variable | Estimate [95% CI] | $\chi^2$-value | p-value |
| ECG-HR | 0.06 [-0.11, -0.01] | $\chi^2(1) = 6.18$ | .01** |
| RESP-RMSSD | -0.0002 [-0.0004, 0.0] | $\chi^2(1) = 4.22$ | .04** |
| EDA-Std tonic | 0.56 [-0.23, 1.34] | $\chi^2(1) = 1.86$ | .2 |

Table 6. Summary of Model (d)'s Results.

| Model d: MCDD (Mental Fatigue, Subjective) | | | |
|---|---|---|---|
| Independent Variable | Estimate [95% CI] | $\chi^2$-value | p-value |
| ECG-RMSSD | 0.004 [0.0005, 0.008] | $\chi^2(1) = 5.10$ | .03** |
| EDA-Mean tonic | -0.14 [-0.37, 0.10] | $\chi^2(1) = 1.32$ | .3 |
| EDA-Max tonic | 0.14 [0.01, 0.28] | $\chi^2(1) = 4.14$ | .04** |

Table 7. Summary of Model (e)'s Results.

| Model e: Combined Datasets | | | |
|---|---|---|---|
| Independent Variable | Estimate [95% CI] | $\chi^2$-value | p-value |
| ECG-HR | 0.04 [0.01, 0.07] | $\chi^2(1) = 8.52$ | .004** |
| ECG-RMSSD | -0.0004 [-0.0009, 0.0] | $\chi^2(1) = 4.22$ | .04** |
| RESP-Mean Amplitude | -0.0001 [-0.0001, 0.0] | $\chi^2(1) = 8.21$ | .004** |
| EDA-Min tonic | -0.0001 [-0.0003, 0.0] | $\chi^2(1) = 6.86$ | .009** |
| EDA-SCR Mean Amplitude | -0.13 [-0.24, -0.02] | $\chi^2(1) = 5.48$ | .02** |
| EDA-SCR Peak Num | 0.01 [-0.003, 0.02] | $\chi^2(1) = 2.31$ | .13 |
| Drowsy Type | - | $\chi^2(3) = 27.56$ | <.0001** |

We first identified key factors associated with different drowsiness types, but with the same type of drowsiness assessment. As shown in Table 3, 4. Features extracted from ECG, RESP, and EDA, including ECG-HR, RESP-RR, EDA-Min tonic, were linked to both physical fatigue and low arousal drowsiness. However, their directions of association may vary. For example, we notice that physical fatigue and low arousal drowsiness led to opposite changes in RESP-RR and EDA-Min tonic. At the same time, drowsiness from sleep deprivation (Table 5) and mental fatigue (Table 6) were both associated with ECG and EDA indicators. Finally, while the RESP-RMSSD was associated with the drowsiness caused by sleep deprivation, no RESP-related feature was associated with drowsiness induced by mental fatigue.

Next, Table 7 further shows that certain features from all three physiological measures were significantly associated with levels of drowsiness, regardless of the types of inducers. Through post-hoc analysis, we found that the drowsiness induced by mental fatigue had lower likelihood of being identified as drowsy compared to those induced by physical fatigue (Odds Ratio [OR] = 0.04, 95% confidence interval [95%CI]: [0.01, 0.16], $\chi^2(1)$ = 21.30, $p$ < .0001), sleep deprivation (OR = 0.19, 95%CI: [0.05, 0.63], $\chi^2(1)$ = 7.26, $p$ = .007), low arousal (OR = 0.04, 95%CI: [0.01, 0.15], $\chi^2(1)$ = 20.57, $p$ < .0001). Besides, drivers induced by low arousal (OR = 5.23, 95%CI: [1.23, 22.32], $\chi^2(1)$ = 5.00, $p$ = .03) and physical fatigue (OR = 4.65, 95%CI: [1.25, 17.33], $\chi^2(1)$ = 5.24, $p$ = .02) tended to be more likely to be identified as drowsy compared to sleep deprivation. Furthermore, we also noticed that, compared to subjective assessments, objective assessments were more likely to identify participants as drowsy (OR = 12.16, 95%CI: [3.28, 45.12], $\chi^2(1)$ = 13.96, $p$ = .0002).

## DISCUSSION

To answer the three research questions, we investigated four datasets regarding four different drowsiness/fatigue types and two drowsiness assessment types.

First, for RQ1, our results indicate that distinct inducers of drowsiness can elicit different physiological variations. We identified that physical fatigue was associated with stronger respiratory and electrical skin activities, likely due to high sympathetic nervous system activation and increased energy expenditure when experiencing physical fatigue (Posada–Quintero et al., 2024), whereas low arousal drowsiness reflects a state of reduced activation in the body. In contrast, sleep deprivation and mental fatigue were primarily associated with cardiac metrics, indicating that central nervous system fatigue predominantly modulates heart rate dynamics (Bourdillon et al., 2021). Notably, post-hoc comparisons showed that mental fatigue was less likely to be detected as drowsy than other inducers (e.g., OR = 0.04 compared to physical fatigue), suggesting that data-driven models trained on one inducer may underperform when applied to another. These findings underscore the necessity of including diverse drowsiness inducers in training datasets to capture the full spectrum of physiological responses.

Further, for RQ2, a comparison of subjective and objective drowsiness ratings showed that objective assessments (such as independent rater annotations and behavioral markers) had significantly higher sensitivity in detecting drowsiness (OR = 12.16), although this may come at the expense of specificity. This increased detection capability suggests that objective measures can capture subtle physiological changes that subjective reports might miss. However, relying solely on objective labels could lead to a higher rate of false positives in real-world situations. Therefore, using both subjective scales (e.g., KSS) and objective indicators (e.g., (Wierwille & Ellsworth, 1994)) can offer a more balanced approach, allowing for adjustments to balance both sensitivity and specificity of the model. Future studies should measure the trade-offs between miss rates and false alarm rates to identify the more effective strategies for combining these labels.

Lastly, for RQ3, we found that, despite inducer- and assessment-specific variations, variations of several features emerged as reliable across all datasets, including a higher heart rate but more stable cardiac activity, reduced respiratory amplitude, and lower skin electrical activation. These cross-condition markers reflect broad autonomic shifts accompanying impaired vigilance and offer a parsimonious feature set for developing generalized detection algorithms. By focusing on these robust indicators, researchers can streamline model complexity and improve external validity when deploying systems across diverse driving scenarios and populations.

## LIMITATIONS

This study relied on existing datasets, which limits participant diversity and standardization of protocols. Further, no public dataset encompasses drowsy driving induced by physical fatigue and thus we adopted a non-driving dataset, which should be addressed in future work.

Second, all datasets were collected in a simulated environment. Future naturalistic driving studies are needed to validate the generalizability of these physiological features.

## CONCLUSION

In this study, we extensively examined the key physiological features of drowsiness, which were induced by various inducers and assessed through different methods. We observed significant differences in the physiological features when different inducers induced drowsiness. This explains why training data-driven models on a single dataset may lead to unsatisfactory performance when the model is used for the detection of other types of drowsiness states. In addition, we further evaluated the diverse effects of drowsiness-inducing inducers, which can provide insights into the drowsiness induction protocols in future experiments. Since we observed that objective ratings have higher sensitivity to drowsiness compared to subjective ratings, we suggest that future drowsiness detection models and assessment protocols incorporate both subjective and objective metrics to balance the miss and false alarm rates. Finally, we also identified common physiological features (i.e., ECG, RESP, and EDA) across multiple datasets, which provided a basis for generalizable model development. We thus call for future drowsiness detection research to pay more attention to the evaluation of model generalizability.